\documentclass{ws-ijmpe}

\begin{document}

\markboth{Authors' Names}{Instructions for  
Typing Manuscripts (Paper's Title)}

\catchline{}{}{}{}{}

\title{MONOPOLE SHIFT IN ODD NEUTRON-RICH F ISOTOPES: A SHELL MODEL DESCRIPTION}

\author{P.C.~SRIVASTAVA\footnote{Present address: Physical Research Laboratory, Ahmedabad-380 009, India.}~ and I.~MEHROTRA}

\address{ Department of Physics, University of Allahabad, Allahabad-211002, India}

\maketitle

\begin{history}
\received{(received date)}
\revised{(revised date)}
\end{history}

\begin{abstract}
        Monopole shift in the proton states in odd neutron rich F isotopes has been studied with neutron number changing from N = 10 to 20 for three different interactions. The variation of monopole shift point out towards appearance of shell closure at N = 14 and N = 16 and its disappearance at N = 20. Low lying states of odd F isotopes have also been calculated in the limited configuration space of {\it sd} shells for these interactions.
\end{abstract}

\maketitle

\section{Introduction}
\label{s_intro}
 Nuclear shell model was started by Mayer and Jensen in 1949 by identifying the magic numbers and their origin.\cite{May49,Hax49,May55} Since then the study of nuclear structure has been carried out on the basis of shell structure associated with the magic numbers. However for a long time the study has remained confined to stable nuclei which lie on or close to the line of $\beta$ stability in the nuclear chart. Present day experimental research in nuclear structure has made extensive use of radioactive ion beams (RIB) in order to study nuclear properties farther away from the region of $\beta$ stability. Detailed experimental data on the properties of very unstable nuclei with neutron/proton ratios radically different from those of stable nuclei has been obtained. Such exotic nuclei exhibit some novel phenomenon like halo structure and modification of shell closure. The shell structure of nuclei with large proton or neutron excess seems to be quite different from that in $\beta$ stable nuclei. It has been demonstrated in earlier studies that neutrons in neutron rich nuclei experience a mean field with much more diffused boundary and resulting single particle energies tend to be more uniformly distributed. It is of interesting to study the changes in the mean field in going from stable to exotic nuclei and to trace its origin. Ideal candidates for such studies are nuclei near doubly magic system with an extra proton or neutron outside the closed core. One such example is neutron rich F isotope having one extra proton outside the $^{16}$O core and a number of valence neutrons. Experimental data on neutron rich F isotopes with N = 9$-$20 has also been made available recently.\cite{NNDC,Mic06}

        In the present work we have studied the evolution of proton mean field for neutron rich F isotopes in going from N = 9 to N = 20 in the shell model frame work. The variation in the single particle energy of the odd proton as a function of changing number of valence neutrons is called the monopole energy shift.

        The paper is organized as follows: In Section 2 theoretical formalism and different types of pn interactions used in the calculation are described. In Section 3 results and discussion are presented. Finally in Section 4 we give conclusions.

\section{ Theoretical formalism}
\subsection{ Monopole shift}
            In F isotopes a valence proton in {$\it j_\pi$} orbital interacts with a number of valence neutron particles filling a set of neutron orbitals {$\it j_\nu$}.\\
The Hamiltonian can be written as
\begin{equation}
       \hat{H}= \epsilon_{j_\pi} + \sum_{j\nu} \epsilon_{j_\nu} \hat{n}_{j_\nu} + V_{\pi\nu}
  \end{equation} 
where $\epsilon_{j_\pi}$ and $\epsilon_{j_\nu}$ denotes the single particle energies of proton and neutron corresponding to states j$_\pi$ and j$_\nu$ respectively, n$_{j_\nu}$ is the number operator for neutrons occupying state j$_\nu$ and V$_{\pi\nu}$ represents the residual proton-neutron interaction. The neutron-neutron interactions within the valence neutron space have been neglected.

If V$_{\pi\nu}$ interaction is decomposed into different multipoles, $\hat{H}$ can be written as
\begin{equation}
       \hat{H}= \hat{H}_{\lambda=0} + higher~ \lambda~ multiples~ of~ \pi\nu ~interaction
  \end{equation} 

where the monopole Hamiltonian $\hat{H}_{\lambda=0}$ is given by
\begin{equation}
       \hat{H}_{\lambda=0}=  \sum_{j\pi} \tilde {\epsilon}_{j_\pi} \hat{n}_{j_\pi} + \sum_{j\nu} \tilde {\epsilon}_{j_\nu} \hat{n}_{j_\nu}  
  \end{equation} 
and $ \tilde{\epsilon}_{j_\pi}$ is the monopole corrected single particle proton energy which is given by

\begin{equation}
\tilde {\epsilon}_{j_\pi} = \epsilon_{j_\pi} + \sum_{j\nu} \bar{E}(j_\pi j_\nu)\hat{n}_{j_\nu}
\end{equation}

The operator $\hat{n}_{j_\nu}$ in Eq.(4) is replaced by its expectation value $\langle \hat{n}_{j_\nu} \rangle$= $\hat{n}_{j_\nu}$, the number of neutrons occupying orbital j$_\nu$.\cite{Sha63,Hey94,Law80,Bru77,Duf96,Smi04}

The term $\bar{E}(j_\pi j_\nu)$ is angular momentum averaged interaction energy

\begin{equation}
 \bar{E}(j_\pi j_\nu)= \frac{\displaystyle{\sum_{J}(2J+1)\langle j_\nu j_\pi;J\mid V \mid j_\nu j_\pi;J\rangle}}{\displaystyle{\sum_{J}(2J+1)}}
\end{equation}

The modified single particle energies $\tilde {\epsilon}_{j_\pi}$ in Eq.(4) are also called the effective single particle energies (ESPE).\cite{Ots01} By construction of the monopole Hamiltonian these ESPE correspond to the spherical shell structure. 

        Thus the single particle energy of one proton in {$\it j_\pi$} is modified when neutron particles are being added to the valence orbital {$\it j_\nu$} just outside the core.
 \subsection{ Interaction}
      In the present work three different neutron-proton interactions have been considered: modified surface delta interaction (MSDI) and effective interactions obtained for ${\it sd}$ shell \cite{Bro06,Bro88} referred in the literature as USDA and USDB. MSDI \cite{Wil71} is a zero range interaction. In USD interaction two body matrix elements have been fitted so as to reproduce the spectrum of stable nuclei in the ${\it sd}$  shell region. In the present work model space for all F isotopes consists of 0d$_{5/2}$, 1s$_{1/2}$  and 0d$_{3/2}$ orbitals for both the protons and neutrons. The ESPE of proton for 0d$_{5/2}$, 1s$_{1/2}$  and 0d$_{3/2}$ orbitals has been estimated from Eqs. (3) and (4) for the above three different interactions.
\begin{table}[h]
\begin{center}
\caption{
 The proton single particle energies for different interactions.} 
\begin{tabular}{cccc}
\hline
Interaction  & d$_{5/2}$ & s$_{1/2}$ & d$_{3/2}$  \\		   
\hline
MSDI  & -7.5600000 & -6.030000 & -3.960000  \\
USDA  &  -3.9436000 & -3.061200 & 1.979800 \\
USDB  &   1.9798000 & -3.207900 & 2.111700  \\ 
\hline           
\end{tabular}
\end{center}
\end{table}
\section{ Results and Discussion}
  The experimental energy levels of low lying positive parity isotopes of F are shown in Fig.1. The ground state of odd F isotopes from N = 12$-$20 is 5/2$^+$ and can be well described by the proton in the lowest configuration of 0d$_{5/2}$ and neutrons paired to J = 0$^+$. The first and the second excited states with J = 1/2$^+$ and 3/2$^+$ respectively also confirm to this configuration. The relative spacing of 1/2$^+$ and 3/2$^+$ from the ground state in $^{21}$F and  $^{23}$F  are indication of monopole shift in the proton energy levels.
\begin{figure}[h]
\begin{center}
\includegraphics[width=7cm]{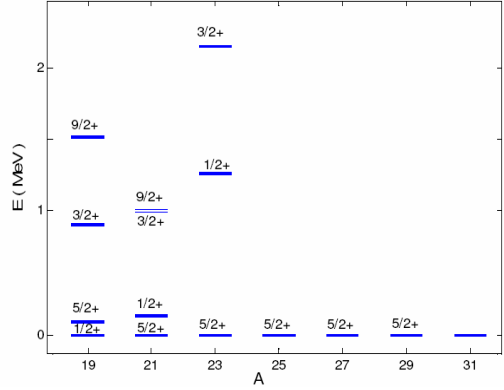}
\caption{
Experimental low-lying states in odd-A F isotopes.}
\label{f_Exf}
\end{center}
\end{figure} 
                                                
        The effective proton single-particle energies  $\tilde {\epsilon}_{j_\pi}$ for F isotopes have been calculated from Eq.(4) for four different types of interactions. The proton single-particle energies $\epsilon_{j_\pi}$ are given in Table 1 for three different interactions. The results of our calculations are given in Figs 2-4 showing 1s$_{1/2}$ and od$_{3/2}$ energies relative to 0d$_{5/2}$ state for different interactions as a function of neutron number. It is observed that for USDA and USDB interaction 1s$_{1/2}$ level rises relative to 0d$_{5/2}$ state reaching a maximum at N = 14 with ε1s$_{1/2}$-ε0d$_{5/2}$= 6 MeV. With further increase in N, 1s$_{1/2}$ state dips in energy coming close to 0d$_{5/2}$ state reaching its minimum at N = 16 and then again it rises. The ε0d$_{3/2}$ - ε1s$_{1/2}$$\cong$5 MeV at N = 16. Similar trend is observed for MSDI interaction though the gaps are much smaller. The 1s$_{1/2}$ level lies midway between 0d$_{5/2}$ and 0d$_{3/2}$ at N = 20 for all the interactions. These features are indicative of appearance of shell closure at N = 14 and N = 16 and its disappearance at N = 20.

\begin{figure}
\begin{center}
\includegraphics[width=7cm]{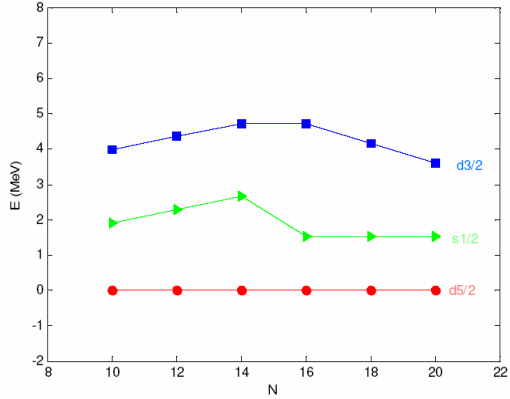}
\caption{
(Color online) Monopole shift in fluorine isotope for the modified surface delta interaction [MSDI].}
\label{f_mmsdi}
\end{center}
\end{figure}
\begin{figure}
\begin{center}
\includegraphics[width=7cm]{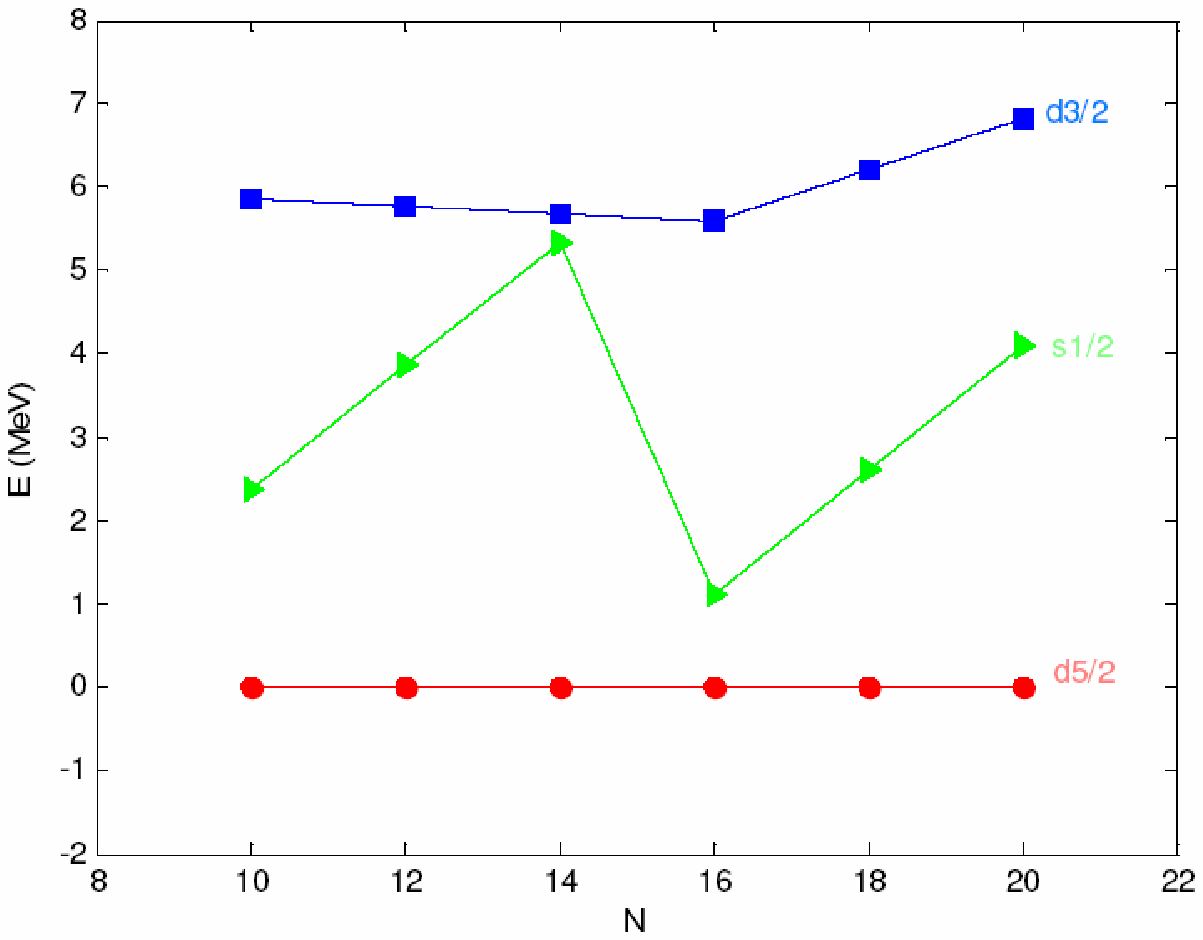}
\caption{
(Color online) Monopole shift in fluorine isotope for the USDA interaction.}
\label{f_mmsda}
\end{center}
\end{figure}
\begin{figure}
\begin{center}
\includegraphics[width=7cm]{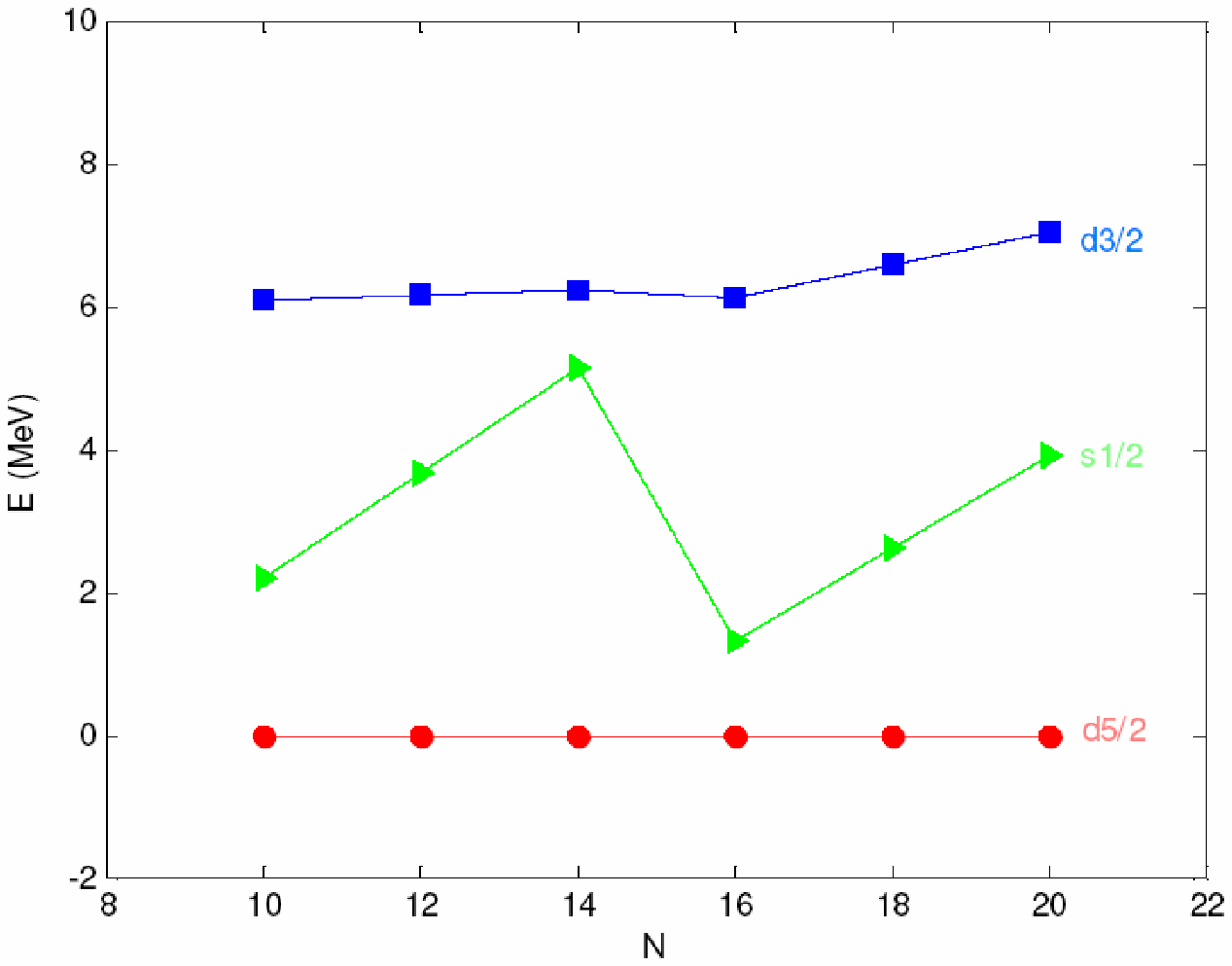}
\caption{
(Color online) Monopole shift in fluorine isotope for the USDB interaction.}
\label{f_mmsdb}
\end{center}
\end{figure}
            
\begin{figure}
\begin{center}
\includegraphics[width=7cm]{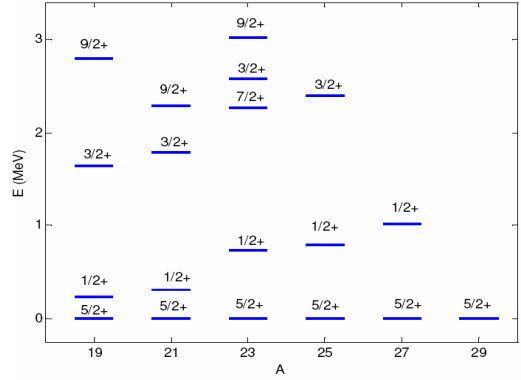}
\caption{
Calculated energy levels in odd-A F isotopes for the delta interaction [MSDI].}
\label{f_Emsdi}
\end{center}
\end{figure}
\begin{figure}
\begin{center}
\includegraphics[width=7cm]{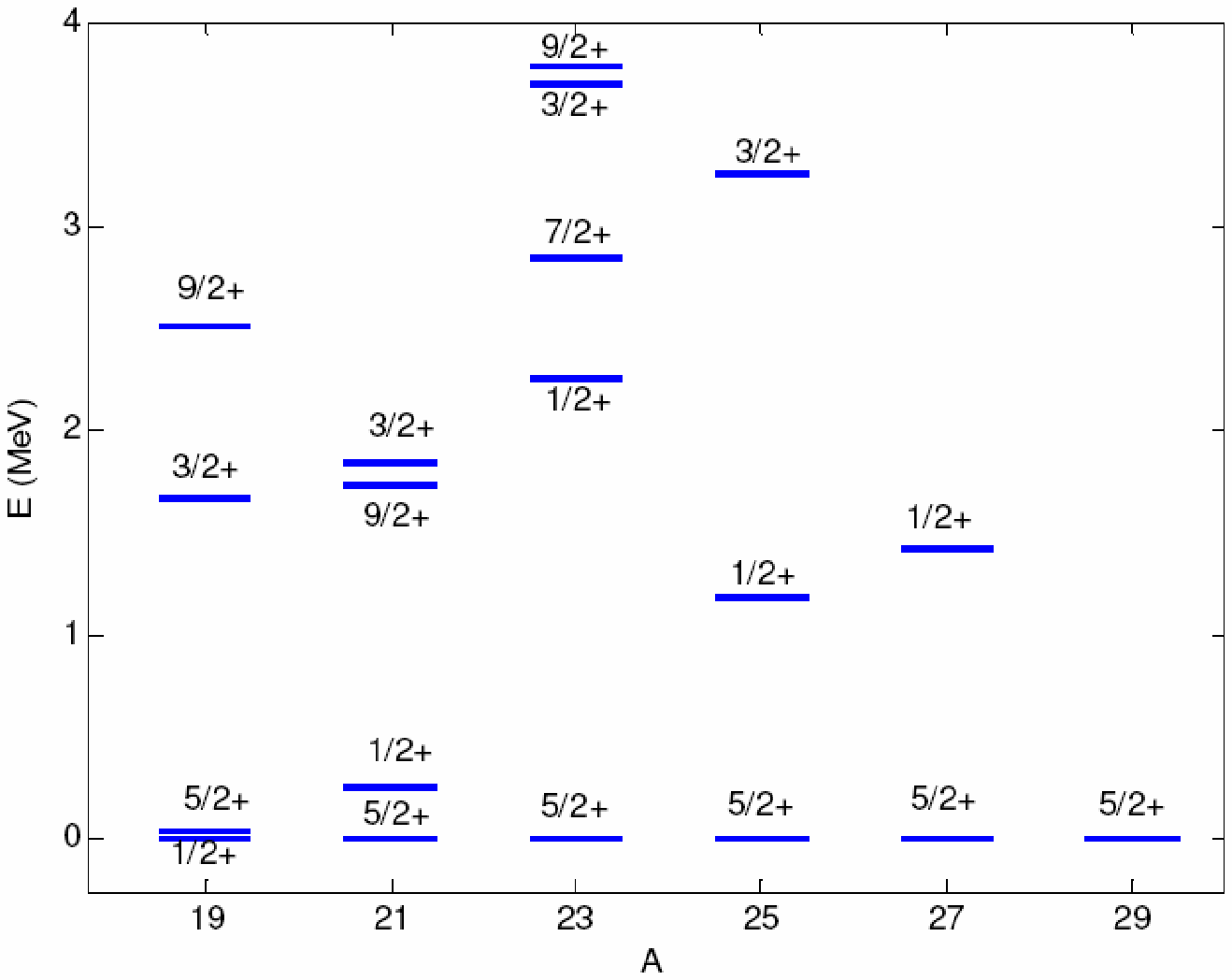}
\caption{
Calculated energy levels in odd-A F isotopes for the USDA interaction.}
\label{f_Eusda}
\end{center}
\end{figure}
\begin{figure}
\begin{center}
\includegraphics[width=7cm]{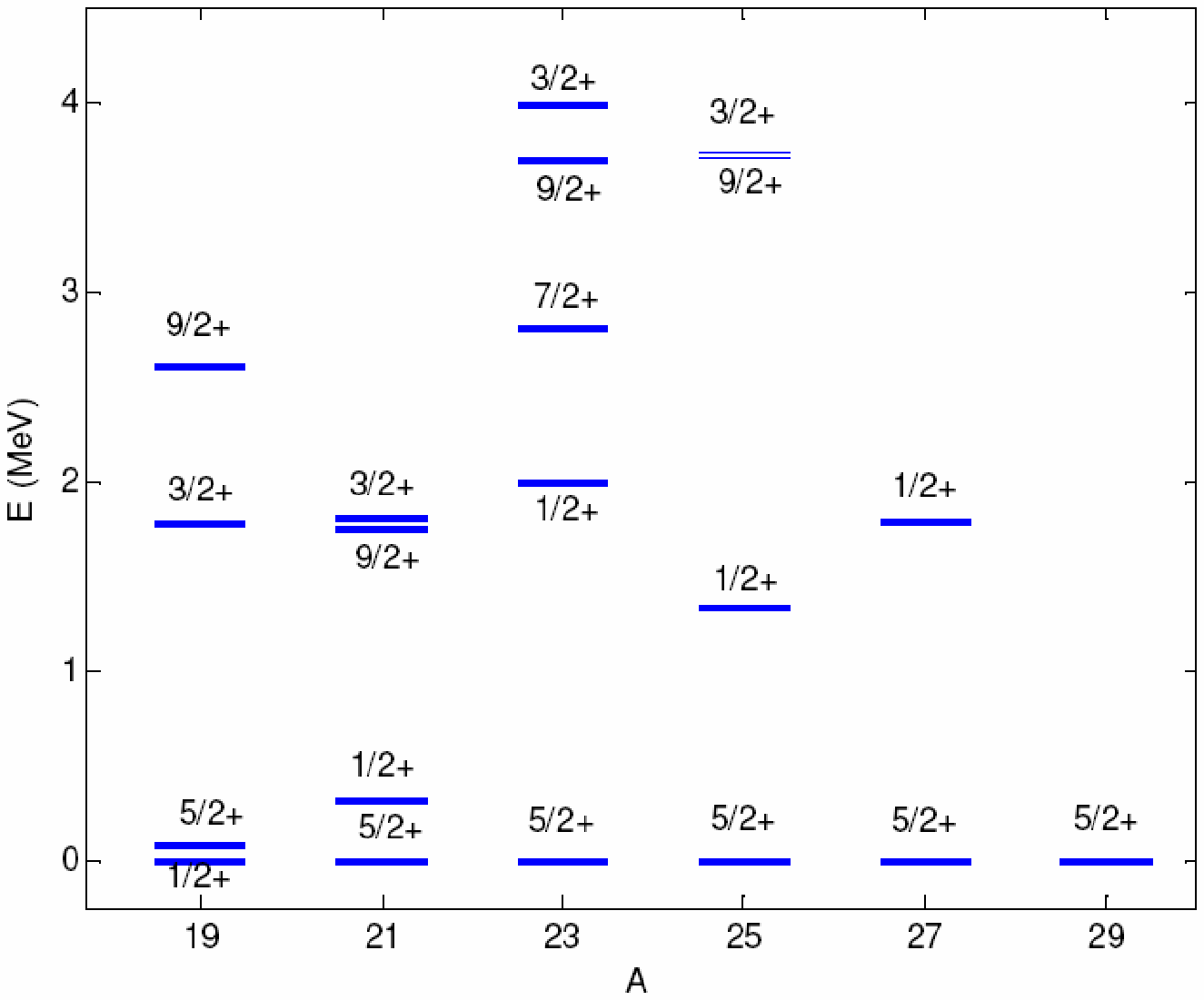}
\caption{
Calculated energy levels in odd-A F isotopes for the USDB interaction.}
\label{f_Eusdb}
\end{center}
\end{figure}

 Low lying energy levels of $^{19-29}$F isotopes have also been calculated with all the three interactions using shell model code Nushell.\cite{Bro07} The shell model space chosen comprises of 0d$_{5/2}$, 1s$_{1/2}$ and 0d$_{3/2}$ for both protons and neutrons. The results obtained for MSDI interactions are shown in Fig. 5. With USDA interaction the ground state spin and correct ordering of the levels reproduced for all the isotopes which are shown in Fig. 6. Similar results are obtained for USDB which are shown in Fig. 7. One important feature of the USDA spectrum is that the first excited state lies high in energy at N=14 and N=16 a characteristic feature of closed shell nuclei. MSDI interaction, predict ground state spin 5/2$^+$ for  $^{19}$F in contrast to the experimental value of 1/2$^+$.

\newpage
\section{Conclusions}
The monopole energy shift for the proton orbital in $^{19-29}$F isotopes have been studied as a function of changing neutron number (N) for three different  interactions. The variation of ESPEs point out towards appearance of shell closure at N = 14 and 16 and its disappearance at N = 20. In writing the monopole Hamiltonian the average matrix element for proton-neutron interaction have been weighted with the number of neutrons occupying that particular orbital. This is consistent with and Hartree Fock approach in which an orbital is either empty or filled with a given occupation probability. Therefore the monopole Hamiltonian is the natural tool to study the mean field through the varying single particle energies connected to mean field. It is also important to compare the ESPEs with experimental centroid energies but in the case of neutron rich fluorine isotopes the experimental data are very sparse.\\
        The strength of the above method can be appreciated if large space shell model calculations are performed diagonalizing the full Hamiltonian $H = H_ {\lambda = 0} + {\displaystyle \sum_{\lambda} H_{\lambda \ne 0}}$ using the same effective two body interactions and comparing the results obtained in two approaches. Large scale shell model calculations are particularly important in deciding about the disappearance of shell closure at N = 20, where the 0d$_{3/2}$ level should lie closer to {\it fp} shell. Such calculations are on the way and shall be reported shortly.

\section*{Acknowledgements}
  Discussions with P Van Isacker on monopole shift is gratefully acknowledged. This work has been supported by UGC, India.

\end{document}